\documentclass[conference]{IEEEtran}
 
\usepackage{cite}
\usepackage{amsmath,amssymb,amsfonts}
\usepackage{graphicx}
\usepackage{textcomp}
\usepackage{xcolor}
\usepackage{url}
\usepackage{multirow}
\usepackage{tabularx}
\usepackage{comment}
\usepackage{booktabs}
\usepackage{soul}
\usepackage[noend]{algpseudocode}

\renewcommand{\baselinestretch}{0.98}
\def\BibTeX{{\rm B\kern-.05em{\sc i\kern-.025em b}\kern-.08em
    T\kern-.1667em\lower.7ex\hbox{E}\kern-.125emX}}

\begin{document}

\title{Scalable Microservice Forensics and Stability Assessment Using Variational Autoencoders}

\author{\IEEEauthorblockN{Prakhar Sharma, Phillip Porras, Steven Cheung, James Carpenter, Vinod Yegneswaran} {\em Computer Science Laboratory, SRI International}}

\maketitle

\begin{abstract}

We present a deep-learning-based approach to containerized
application runtime {\em stability analysis}, and an intelligent publishing algorithm that can dynamically adjust the depth of process-level forensics published to a backend incident analysis repository.  The approach applies variational autoencoders (VAEs) to learn the stable runtime patterns of container images, and then instantiates these container-specific VAEs to implement stability
detection and adaptive forensics publishing. In performance comparisons using a 50-instance
container workload, a VAE-optimized service versus a conventional eBPF-based forensic publisher demonstrates 2 orders of magnitude (OM) CPU performance improvement, a 3 OM reduction in network transport volume, and a 4 OM reduction in Elasticsearch storage costs. We evaluate the VAE-based stability detection technique against two attacks, CPUMiner and HTTP-flood attack, finding that it is effective in isolating both anomalies.  We believe this technique provides a novel approach to integrating fine-grained process monitoring and digital-forensic services into large container ecosystems that today simply cannot be monitored by conventional techniques.
\end{abstract}


\section{Introduction}\label{intro}

Application containerization is becoming the dominant technology to
architect and deploy microservice applications and software
services. IDC Research predicts that by 2023 more than 500
million applications and services will be developed using cloud-native
containerization services ~\cite{IDC-Report}, which is roughly the same number of
applications that have been developed over the last 40
years combined. Containerization has become
integral to the security and management of front-line software
services and sensitive data processing applications across all major
industries in our economy.

However, container-based microservice ecosystems pose a myriad of challenges to the
integration of security services, runtime management, and forensics
collection for security incident response; most
significant of which is scaling these services to manage modern container pipelines.  Process monitoring and forensic logging impose significant costs to container workload
overhead, as they consume {\em CPU cycles,} {\em network bandwidth,} and {\em data storage}, to deliver and store information to log repositories. In fact, 
industrial scale container ecosystems (thousands of 
application instances and beyond) are largely out of reach of
conventional process-level monitoring and forensic services. Nevertheless, fine-grained process
monitoring remains an important service, and indeed a compliance
requirement, for enabling security, fault management, and
damage assessment.

This paper examines two critical questions pertaining to the
secure runtime management of such large-scale container ecosystems.
First, {\em how can we conduct massively scalable monitoring of
containerized applications to identify when containers become unstable
or subverted}?  Second, {\em when runtime instabilities are perceived,
how can we capture enough fine-grained digital forensics to drive an
incident response or fault analysis}?

For container DEVOPS environments, process visibility is commonly
achieved through some form of extended Berkeley Packet Filter
(eBPF)-based tracing, such as through services like Sysdig\cite{Sysdig} and
Chisel\cite{Chisel}.  These tools offer a significant performance advantage
over typical host auditing services for process system-call logging.
eBPF extends a host kernel with logic to trap key system calls on entry or exit from the kernel, providing a fine-grained view of process-internal runtime behavior to
drive intrusion detection, fault analysis, and other forms of
anomaly detection. 

We propose a new strategy for conducting a computationally-efficient local analysis of container runtime behavior. This is done through an extraction of
{\em activity vectors} that captures statistics and attributes from a container's eBPF-based forensics produced during intervals of runtime operation.  An activity vector is input to a VAE that determines the extent to which the vector fits the container's trained {\em stable baseline} runtime model. 
VAEs are robust and efficient in performing event-based pattern analysis ~\cite{seasonalKPIs,zavrak}. 
We introduce a VAE-driven adaptive forensic publisher, which produces a runtime 
stability assessment to determine the depth of forensics published to the remote 
log repository. During an interval of runtime stability, the VAE-based forensic 
publisher delivers an {\em activity model} that captures the modality of the 
application as it was observed through its eBPF-derived system-call forensics.     
An activity model is a highly reduced representation of the forensic data, requiring a 
fraction of transport and storage cost. During intervals not matching the 
trained runtime pattern, the publisher delivers the interval's forensic 
log along with the activity model.  Analytics are then performed on the forensic 
logs to isolate evidence of errors or faults, as well system-call actions that 
may indicate malicious operations performed during the unstable runtime interval.

We present an overview of the VAE algorithm for interval-based
container stability assessment and our adaptive forensic publisher
service. The system has been implemented and tested within Docker and
Kubernetes environments. Results of a comparative
performance assessment of the VAE-based forensic publisher versus
conventional process-level forensic collection services are presented. These results
demonstrate the potential for deep-learning optimized forensic
publishing services to enable fine-grained process monitoring and
forensic-based incident response service to scale within ecosystems
that today are beyond the resource consumption limitations of conventional
monitoring techniques.
 
\begin{figure}[t]
\begin{center} 
    \includegraphics[scale=0.65]{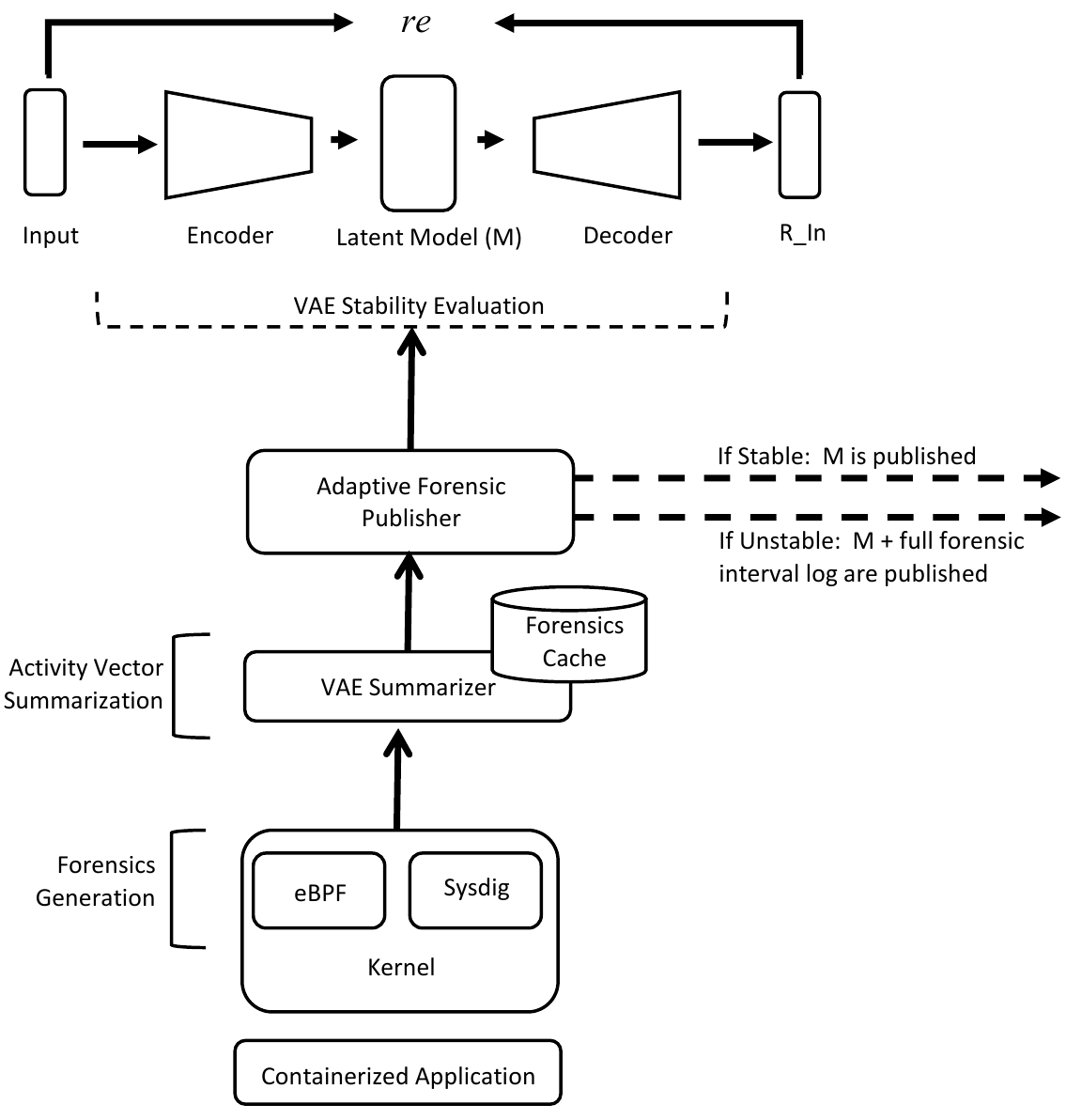}\\
    Fig 1. Design overview of VAE-based adaptive 
    forensic publishing service for containerized applications.
    \label{f:vae-arch}
\end{center}
\end{figure}

\section{VAE Design}\label{design}
Given any dataset $X$ drawn i.i.d  from true data distribution $p_{data}$, variational auto-encoders \cite{b12} (VAEs) aim to learn an encoder decoder pair that enable drawing new samples $x_{new}$ such that $x_{new} \sim p_{data}$ ($\sim$ signifies a sample drawn from a probability distribution). The encoder and decoders are usually represented by neural networks. The generative process of VAEs involves learning a low dimensional latent space and is defined as: 
$$z_{new} \sim p(Z)\ \ \ x_{new} \sim p_\theta(X|Z=z_{new})$$ 

where $p(Z)$ is a fixed prior distribution over latent space $Z$. VAE operation can be broken down into two parts, first the input is mapped to the latent space via an encoder $E$ (parametrized by $\phi$) 

$$E_\phi(x)=z \sim q_\phi(z|x) = q(Z|f_\phi(x))$$ 
\vspace{-0.1in}

and then the latent space is mapped to the original distribution via a decoder $D$ (parametrized by $\theta$) 
\vspace{-0.1in}

$$D_\theta(z) = x_{new} \sim p_\theta(x|z) = p(X|g_\theta(z))$$ 


Here $f_\phi(x)$ and $g_\theta(z)$ represent neural networks. The VAE architecture is trained via gradient descent by minimizing the evidence lower bound (ELBO) for each sample $x$.


$$log\ p_\theta(x) \geq ELBO(\phi, \theta, x)$$ 

\vspace{-0.2in}

$$=\mathbf{E}_{z \sim q_\phi(z|x)} log\ p_\theta(x|z) - KL(q_\phi(z|x)||p(z))$$ 



Abstractly, the elements that compose a typical VAE are illustrated at the top of Figure 1. 
The term $\mathbf{E}_{z \sim q_\phi(z|x)} logp_\theta(x|z)$ is called the reconstruction error or reconstruction loss ($L_{rec}$) of each sample, often calculated in practice using the mean squared error (MSE) between the data sample and it's reconstruction. For a VAE trained on $X \sim p_{data}$, $L_{rec}$ would be high if the sample is not drawn from $p_{data}$ and would be low for a sample drawn from $p_{data}$. Post training, $L_{rec}$ for any new data sample can be used as an indicator of how well the new sample conforms to the original data distribution from which the VAE was trained. Thus, VAEs have been used for anomaly detection tasks in recent years ~\cite{seasonalKPIs,betaVAE,reconprob}. 

Figure 1 also illustrates the major processing layers in our
implementation of an eBPF-based
container forensic publishing service that employs 
VAE-based stability assessment.  


\vspace{0.1in}

\noindent \textbf{Forensics Generation: } 
The primary source of process forensics are captured via eBPF and Sysdig,  
configured to generate forensic records for a selected set of
system calls.  Our configuration tracks a default set of 72 Linux 
system calls, which represent the most commonly audited system calls \cite{eXpert-BSM} \cite{ASAX} to 
drive security incident analysis, Host IDS analytics, and application
fault analysis. These systems calls can be categorized into 10
classes of activity: process events, the set-user-id family of
events, network events, file and directory access events,
kernel module load events, process and application
virtualization event, file descriptor replication, file
attribute events, filesystem mount events, and IOCTL events.
In addition, the forensic stream can be augmented with sensor
event streams, application log events, system events, and 
system resource metrics.

\vspace{0.1in}

\noindent \textbf{Activity Vector Summarization: } 
At regular temporal intervals, a forensic summarization service, or
{\em summarizer}, computes a vector of statistics from an
analysis of the forensic stream produced during the interval.  These statistics
include continuous and categorical measures of events (or even N-grams), invocation counts, response codes, and other arguments extracted from the
forensic data produced during the interval.  The temporal window is
configurable, with a default value of 30 seconds.  The activity vector 
provides a concise  salient behavioral summarization of the runtime pattern
observed, and used by the VAE to assess each runtime interval against its
trained data model (derived from a set of activity vectors produced during
the application's training period).   

\vspace{0.1in}

\noindent \textbf{Adaptive Forensic Publisher: } 
We introduce an adaptive forensic publisher that alters the 
depth of forensics published based on the VAE's analysis of the 
container activity vector.   Here, the depth
of forensics published to a remote forensic repository is
selected based on whether the container's latest runtime
interval is found to be consistent with the vetted-stable
forensic corpus from which the VAE was trained.  During
intervals in which the VAE-processed forensic stream produced
a low reconstruction error, the VAE latent model M is
published to Elasticsearch.  Otherwise, both M
and the cached forensic stream for this interval are published to Elasticsearch.  The cache, depending on the local
host configuration, can host forensics across multiple prior
intervals, enabling the publisher to serve Elasticsearch with
prior intervals upon request.

\vspace{0.1in}

\noindent \textbf{VAE Stability Evaluation: }With each incoming summarizer message $x_{in}$ per container, the publisher checks whether trained normalization and VAE models for that container exist. They are used to calculate $L_{rec} = MSE(x_{in}, D_\theta(E_\phi(x_{in})))$, which is then compared to a pre-set reconstruction error threshold $r_{th}$. $L_{rec} > r_{th}$ is considered to be a drift in stability, otherwise the operating mode is considered stable.

\vspace{0.1in}

\noindent \textbf{VAE Training: } In case a valid container VAE model does not exist, each incoming vectorized summarizer message is stored until a pre-decided number of total summarizer messages have been accumulated into a dataset. The dataset features are normalized and the VAE is trained on the normalized dataset using mini-batch gradient descent. Post accumulation, both the normalization model and the VAE neural network model are stored to be used for evaluation. Important parameters that govern training and evaluation are explained in the Appendix.
\section{Evaluation}\label{eval}


\subsection{Evaluating Multi-container Workload Overhead}\label{performance}

We conducted a series of comparative assessments of VAE-based publishing overhead to that of a standard eBPF-based container monitoring system. The standard system performs interval caching, and utilizes the Elasticsearch bulk-API transfer protocol per interval, offering a reasonable and performance-aware implementation for typical process forensic publishing. We consider CPU overheard, processing time, network transport overhead, and Elasticsearch data-storage costs for comparison.\\

We used the Falco system-call test-generation container provided by the Falco project~\cite{Falco}.  The experiment employed the use of eBPF and Sysdig, configured to capture 72 Linux system-calls that are commonly used for forensic security analysis and error diagnosis. The experiment was hosted on an Intel(R) Xeon(TM) CPU 3.20GHz, 8 core, 16GB DDR2 DRAM, 2TB SATA II Drive, and Docker hosted on Ubuntu 16. We conducted the performance tests using 10, 20, 30, 40, and 50 container instances on our host, with 30-second caching intervals, and then averaged to performance statistics over 50 intervals.
\begin{figure}[ht]
    \centering
    \includegraphics[width=3.2in]{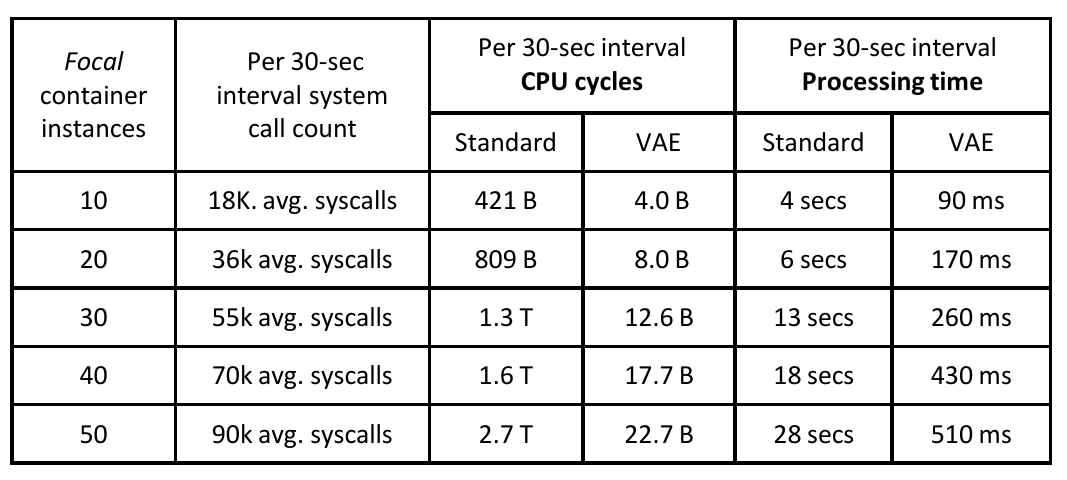}
    \small {\em Table 1. Per interval CPU cycles and processing time for all forensic production and publication computation activity}
    \label{f:cpu-overhead}
\end{figure}

\begin{figure}[ht]
    \centering
    \includegraphics[width=3.2in]{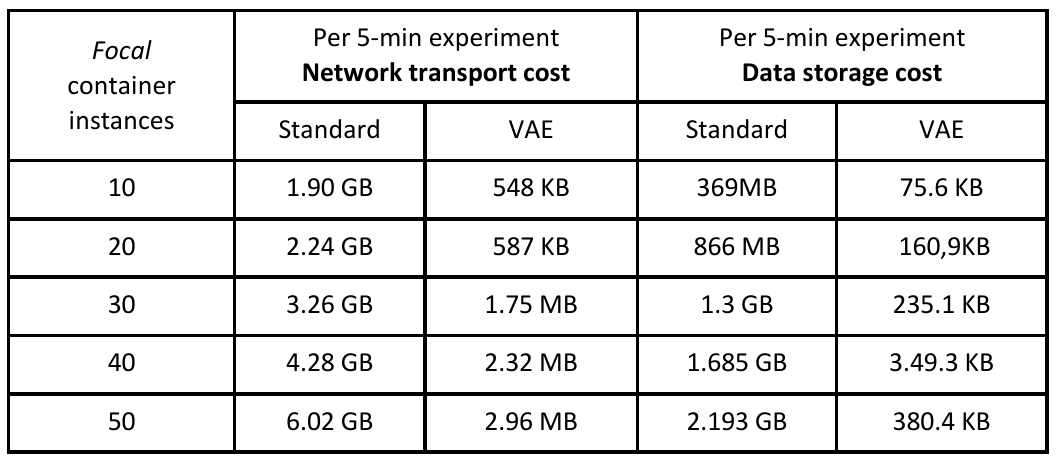}
    \small {\em Table 2. Network transport and Elasticsearch data-storage costs for forensic production and publication}
    \label{f:net-data-overhead}
\end{figure}
\vspace{-2mm}
Table 1 shows the per-interval CPU costs and processing time incurred by the standard publishing method versus the VAE, for each of  five test processing loads.  {\em Perf(1)} (Linux command) was used to compute CPU cycles per the two forensic process algorithms, excluding consideration of the eBPF and sysdig costs, as these two modules are used and configured identically for both publishing systems. Of particular interest is the observation that at 50 container instances (90K average syscalls per interval), the standard publishing method requires approximately 28 seconds to publish the forensics during the average 30 second interval.  In general, when a publisher’s processing time exceeds the collection interval, this is considered to point of saturation, as it takes longer to publish the records than to produce them.  In contrast, under the same 50 container workload, the VAE requires only 510 milliseconds to conduct its stability analysis and publish results.\\

Table 2 shows the network transport overhead and storage costs for the standard publisher versus the VAE. These costs reflect to overhead involved in transferring forensic and container-stability assessments to a forensics data center for incident and fault analysis.  The network transport cost is captured in bytes observed from {\tt tcpdump} captured from the publisher’s communications with Elasticsearch over each 5-minute experiment.  The table also shows the average byte storage increase incurred by Elasticsearch during each 5-minute experiment.  

\subsection{CPUMiner: Detecting Container Hijacking using Microservice Stability Assessment}\label{Stability}

In addition to performance analyses, we conducted stability-veracity assessments to assess whether activity vectors captured during stable operations correspond to a low reconstruction error from a trained model.  Periods in which non-malicious and non-error related operations produce reconstruction errors above threshold represent incompleteness in the training set and, depending on their frequency, may indicate a need for enhanced model retraining. Another assessment perspective is the need to validate that encounters with operations not represented in the trained model will be recognized as unstable intervals by the VAE. Our assessment is applicable to a range of test scenarios, including  application hijacking,  hardware failures, infrastructure faults that impact  the container, network attacks, and other fault and attack related scenarios.

Table 3 illustrates an  assessment test in which an Nginx-trained VAE observes a series of runtime intervals during which a malicious {\em CPUMiner}\cite{CPUMiner} application is integrated into the Nginx container. The assessment begins with intervals of  normal Nginx operation (a shell script that fetches the index page once per second).  As the assessment proceeds, the container experiences a login shell, followed by the issuing of various shell commands. As these actions unfold, the VAE’s reconstruction error grows dramatically, by orders of magnitude as shown in the table. Next,  the VAE observes the downloading and compilation of  CPUMiner, a common application that is installed into hijacked containers to  mine Bitcoins. The VAE produces an astronomical reconstruction error from the initial baseline Nginx operations. Finally, CPUMiner is initiated in the container, continuing the high reconstruction pattern indefinitely, or until CPUMiner no longer pollutes the container’s interval activity vectors. \\

\begin{table}[h]
\begin{center}
\footnotesize
 \begin{tabular}{|c|c|} 
 \hline
 \textbf{Activity} & \textbf{VAE reconstruction error} \\ [0.5ex] 
 \hline
 Normal NGINX processing & 0.014993 \\ 
 \hline
 Connect with bash shell & 6.298320 \\
 \hline
 Execute shell commands & 19390605381.8 \\
 \hline
 CPUminer package download & 1.3524e+13 \\
 \hline
 Compile CPUminer & 13642381656064 \\ 
 \hline
 CPUminer execution & 344375512 \\   \hline
\end{tabular}
\end{center}
\small {\em Table 3. Interval reconstruction errors from a container operating an Nginx-trained VAE to a gradual progression of actions that are used to download CPUMiner and install it within the container}
\label{f:cpuminer}
\end{table}
\vspace{-4mm}
\subsection{HTTP Flooding Attack: Detecting Flood Attacks using Microservice Stability Assessment}\label{httpflood}
\vspace{4mm}
\begin{figure}[ht]
    \begin{center}
        \includegraphics[scale=0.35]{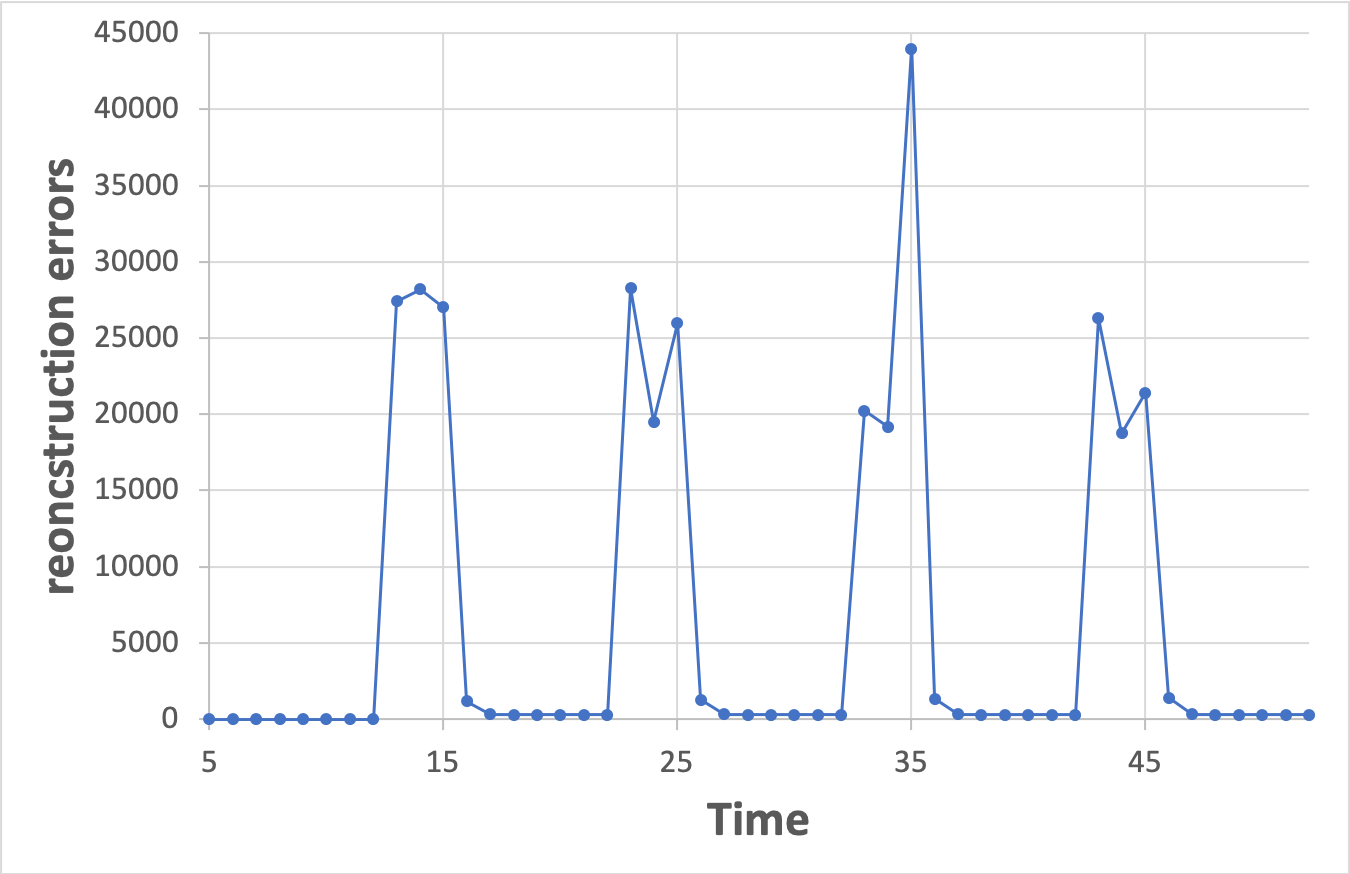}
        \small{Fig 3. Reconstruction errors during a cyclical HTTP-flood attack on the {\tt AtSea\_app} container}
    \end{center}
\end{figure}

We extend the stability-veracity assessment to a Docker-based demo application experiencing an HTTP flood attack. We used {\tt AtSea} \cite{Atsea}, a containerized Java REST application with a database for product inventory, customer data, and orders, a shopping cart, a Nginx reverse proxy implementing HTTPS and a payment gateway to simulate certificate management. We used {\tt Golang-httpflood} \cite{Leeon123} to simulate large number of HTTP requests to overload the {\tt AtSea} containers. An instance of the VAE is trained for the atsea\_app container and the {\tt Golang-httpflood} is invoked cyclically for 1 minute intervals with and 4 minutes of inactivity. Figure 3 shows trained VAE reconstruction errors for such an attack. The spikes correspond to attack durations. 
\section{Related Work}\label{relatedwork}

The topic of anomaly detection in microservices has received recent
research attention.  Ohlsson uses Robust Principal Component
Analysis (RPCA) techniques to detect microservice anomalies~\cite{Ohlsson18}.
Samir et al., describe the use of Markov Models
to detect anomalous container behavior ~\cite{Samir2019}
Naikade uses LSTMs to detect performance anomalies in microservice deployments~\cite{Naikade2020}.
Meng et	al. propose detecting variance in application call trees as
a strategy	to detect microservice performance anomalies~\cite{MENG2021}.  
Magableh uses deep Q-learning networks (DQNs) to build self-adaptive agents 
for performance optimization~\cite{Magableh19}.\\

Other recent works investigate the applicability of deep learning techniques, 
specifically VAEs to the networking domain. Gan et al. \cite{sage} and Pol et al. 
\cite{pol} apply a version of VAEs called conditional VAEs for root cause analysis 
in cloud microservices. Xu et al. employ VAEs for anomaly detection in seasonal 
KPIs of web applications \cite{seasonalKPIs}. For network intrusion detection 
tasks, a case study \cite{casestudy} used vanilla auto encoders, while 
Zavrak et al. used VAEs \cite{zavrak} trained on flow-based features. Akrami et al. 
propose beta-VAEs \cite{betaVAE} and apply them to network data for anomaly 
detection. Other papers have also employed VAE reconstruction probability as the metric 
for anomaly detection instead of the reconstruction error \cite{reconprob}. To 
our knowledge, our work is the first to explore the application of VAEs to 
containerized applications for real-time stability analysis (as opposed to carrying out anomaly detection on benchmark networking datasets, as seen in a majority of the aforementioned works) in conjunction 
with a dynamic forensic publisher.
\section{Conclusion and Future Work}
We present a method to transform container process-level
forensics into interval-based abstract 
activity vectors, and then input these vectors to a VAE
that produces a metric of container runtime stability. This metric is then used to drive an adaptive forensic publisher that alters the depth of information it stores.
The approach enables performant determination of the runtime stability of large numbers of containers and isolation of forensic data for deeper analysis. A comparative analysis of a VAE-optimized publisher versus standard process-level forensic auditing shows that our approach reduces CPU, network, and data storage overhead by orders of magnitude. The results suggest that fine-grained process-level monitoring can be successfully scaled to large container ecosystems that are today outside the resource-consumption costs of traditional
forensic data collection services. Future work includes setting the process-event capturing time window dynamically to make it difficult for an adversary to spread the attack over multiple windows; and extending the list of process level events currently being captured for summarization (details provided in the Appendix).

\newpage
\begin{center}
    \textbf{Appendix}
\end{center}
\textbf{VAE Configuration: }The encoder and decoder modules of the VAE are instantiated each with three hidden layers, each layer has 16 filters in a fully connected fashion. The dimension of the latent model is chosen to be 10. The network is trained with the {\tt Adam} optimizer with learning rate $10^{-4}$, $\beta1 = 0.9$, and $\beta2 = 0.999$. The network is trained for 100 epochs. These hyperparameters were chosen after a grid search on the parameter space for the specific summarizer dataset per-container. The summarizer dataset for each container is normalized using a min-max scaler that is also stored with the VAE model for the purpose of data normalization at test time. 

\vspace{0.1in}

\noindent \textbf{Evaluation Scheme: } We used the  {\tt perf} tool for CPU profiling. Perf stats are captured by attaching the tool to the process IDs of {\tt Sysdig chisel}, {\tt summarizer}, and {\tt publisher} processes running on the Linux test machine. Networking specifics are extracted using a running {\tt tcpdump} process. Storage usage metrics are extracted from elastic search running on a separate server.

\vspace{0.1in}

\noindent \textbf{Reconstruction error and reconstruction threshold: }We employ VAE reconstruction error for each new sample as an indicator of {\em stability}. Throughout training, the VAEs minimize the reconstruction error on a dataset as a means of learning the data distribution. A sample training curve for a VAE trained on a dataset is represented in Fig. 4.
\begin{figure}[h]
\begin{center}
    \includegraphics[scale=0.45]{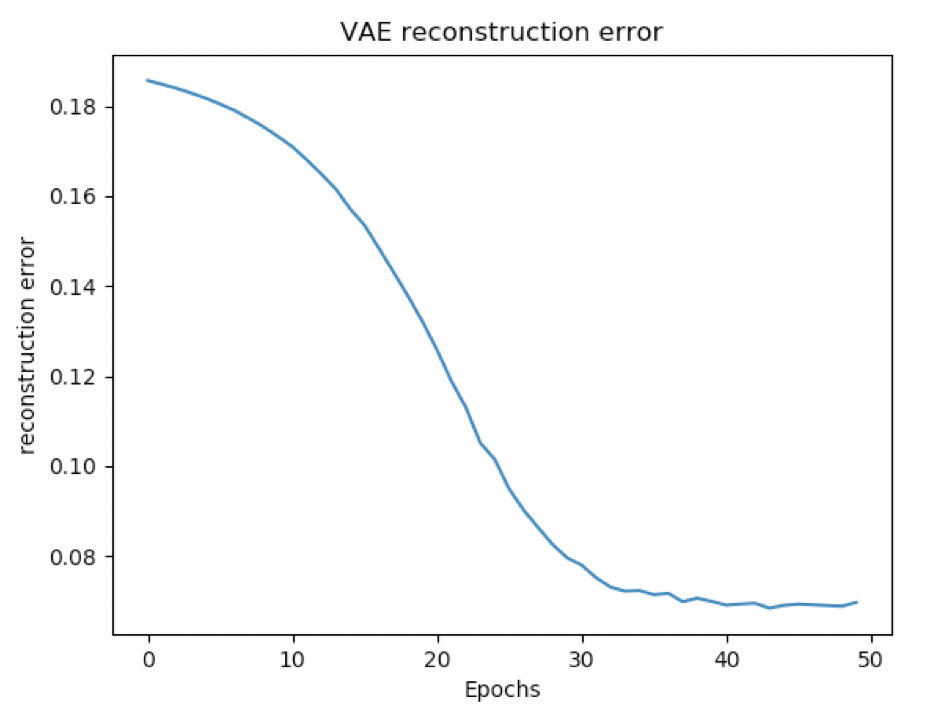}\\
    Fig 4. Reconstruction errors during training. VAE train reconstruction error settles to $r_{last}$ after epoch 40
\end{center} \end{figure}The training reconstruction error settles to a minimum value $r_{last}$ after training. For each new sample, deviation from $r_{last}$ signifies that the data sample is different from the original training data or is {\em out of distribution}. Owing to this fact, we set the reconstruction threshold $r_{th}$ in two major ways for our experiments:
\begin{itemize}
    \item \textbf{Based on a heuristic/experience:} This approach sets $r_{th}$ based on observed reconstruction errors generated during an instable period. \\
    \item \textbf{$k$ standard deviations away from the mean :} This approach records the mean ($r_{mean}$) and standard deviation ($r_{sd}$) of training reconstruction errors in the last training iteration. The reconstruction threshold is then set using the following formula : $r_{th} = r_{mean} + k*r_{sd}$ where $k$ is set by the user. This approach has the benefit of making the VAE more flexible or conservative based on the value of $k$.
\end{itemize}

\vspace{0.1in}

\noindent \textbf{Summarization of Forensic Events: } We used a (non-exhaustive) list of the following kernel events:
\begin{itemize}
\item Process Events:
    \begin{itemize}
        \item common  fork                     \_\_x64\_sys\_fork/ptregs
        \item common  vfork                   \_\_x64\_sys\_vfork/ptregs
        \item 64      execve                       \_\_x64\_sys\_execve/ptregs
        \item 64      execveat                   \_\_x64\_sys\_execveat/ptregs
        \item common  exit                      \_\_x64\_sys\_exit
        \item common  kill                        \_\_x64\_sys\_kill
        \item 64      ptrace                       \_\_x64\_sys\_ptrace
        \item common  prctl                   \_\_x64\_sys\_prctl
        \item common  arch\_prctl         \_\_x64\_sys\_arch\_prctl
        \item x32     execve             \_\_x32\_compat\_sys\_execve/ptregs
        \item x32     ptrace             \_\_x32\_compat\_sys\_execveat/ptregs
    \end{itemize}

\item Set User ID Family Events:
    \begin{itemize}
        \item common  setuid                  \_\_x64\_sys\_setuid
        \item common  setgid                  \_\_x64\_sys\_setgid
        \item common  setpgid                \_\_x64\_sys\_setpgid
        \item common  setsid                   \_\_x64\_sys\_setsid
        \item common  setreuid               \_\_x64\_sys\_setreuid
        \item common  setregid               \_\_x64\_sys\_setregid
        \item common  setgroups            \_\_x64\_sys\_setgroups
        \item common  setresuid             \_\_x64\_sys\_setresuid
        \item common  setresgid             \_\_x64\_sys\_setresgid
        \item common  setfsuid                \_\_x64\_sys\_setfsuid
        \item common  setfsgid                   \_\_x64\_sys\_setfsgid
    \end{itemize}

\item Network Events:
    \begin{itemize}
        \item common  socket                    \_\_x64\_sys\_socket
        \item common  connect                 \_\_x64\_sys\_connect
        \item common  accept                    \_\_x64\_sys\_accept
        \item common  bind                        \_\_x64\_sys\_bind
        \item common  listen                      \_\_x64\_sys\_listen
        \item common  accept4                \_\_x64\_sys\_accept4
    \end{itemize}

\item File and Directory Access Events:
    \begin{itemize}
        \item common  open                            \_\_x64\_sys\_open
        \item common  close                            \_\_x64\_sys\_close
        \item common  openat                      \_\_x64\_sys\_openat
        \item common  mkdir                         \_\_x64\_sys\_mkdir
        \item common  rmdir                          \_\_x64\_sys\_rmdir
        \item common  rename                      \_\_x64\_sys\_rename
        \item common  creat                           \_\_x64\_sys\_creat
        \item common  link                              \_\_x64\_sys\_link
        \item common  unlink                          \_\_x64\_sys\_unlink
        \item common  symlink                       \_\_x64\_sys\_symlink
    \end{itemize}

\item File and Directory Access Events (cont.)
    \begin{itemize}
        \item common  mknod                       \_\_x64\_sys\_mknod
        \item common  mkdirat                     \_\_x64\_sys\_mkdirat
        \item common  mknodat                    \_\_x64\_sys\_mknodat
        \item common  unlinkat                     \_\_x64\_sys\_unlinkat
        \item common  renameat                  \_\_x64\_sys\_renameat
        \item common  linkat                         \_\_x64\_sys\_linkat
        \item common  symlinkat                  \_\_x64\_sys\_symlinkat
        \item common  fchmodat                  \_\_x64\_sys\_fchmodat
        \item common  renameat2               \_\_x64\_sys\_renameat2
    
    \end{itemize}

\item  Kernel Module Load Events:
    \begin{itemize}
        \item 64      create\_module
        \item common  init\_module             \_\_x64\_sys\_init\_module
        \item common  delete\_module           \_\_x64\_sys\_delete\_module
        \item x32     kexec\_load              \_\_x32\_compat\_sys\_kexec\_load
    \end{itemize}

\item Process/App Virtualization Event:
    \begin{itemize}
        \item common clone      \_\_x64\_sys\_clone/ptregs
        \item common clone3     \_\_x64\_sys\_clone3/ptregs
            \item process\_vm\_readv
            \begin{itemize}
                \item 64      \_\_x64\_sys\_process\_vm\_readv
                \item x32   \_\_x32\_compat\_sys\_process\_vm\_readv
            \end{itemize}
            \item process\_vm\_writev
            \begin{itemize}
                \item 64    \_\_x64\_sys\_process\_vm\_writev
                \item x32  \_\_x32\_compat\_sys\_process\_vm\_writev
            \end{itemize}

    
    \end{itemize}

\item File Descriptor Replication:
    \begin{itemize}
        \item common  dup                              \_\_x64\_sys\_dup
        \item common  dup2                            \_\_x64\_sys\_dup2
        \item common  dup3                           \_\_x64\_sys\_dup3
    
    \end{itemize}

\item File Attribute Events:
    \begin{itemize}
        \item common  chmod                      \_\_x64\_sys\_chmod
        \item common  fchmod                     \_\_x64\_sys\_fchmod
        \item common  chown                      \_\_x64\_sys\_chown
        \item common  fchown                     \_\_x64\_sys\_fchown
        \item common  lchown                     \_\_x64\_sys\_lchown
        \item common  fchownat                \_\_x64\_sys\_fchownat
    
    \end{itemize}

\item Filesystem Mount Events:
    \begin{itemize}
        \item common  mount     \_\_x64\_sys\_mount
        \item common  umount2   \_\_x64\_sys\_umount
        \item common  fsmount   \_\_x64\_sys\_fsmount
    
    \end{itemize}

\item IOCTL Events:
    \begin{itemize}
        \item 64      ioctl                     \_\_x64\_sys\_ioctl
        \item x32     ioctl                    \_\_x32\_compat\_sys\_ioctl
    \end{itemize}
\end{itemize}
These event indicators are extracted from the eBPF forensic log and passed through our summarizer module to be distilled in a numerical vector format. A collection of these vectors form our dataset.

\bibliographystyle{IEEEtranS}
\bibliography{9_references, vinod-refs}{}
  
\end{document}